\documentclass{emulateapj}
\shortauthors{Zhong \& Wang}

\newcommand{\ucxb}{2S~0918$-$549}

\begin{document}

\title{The Likely Orbital Period of the Ultracompact Low-Mass X-Ray Binary 2S~0918$-$549} 

\author{Jing Zhong\altaffilmark{1} and Zhongxiang Wang}
\affil{Shanghai Astronomical Observatory, Chinese Academy of Sciences,\\
80 Nandan Road, Shanghai 200030, China}
\email{jzhong@shao.ac.cn, wangzx@shao.ac.cn}

\altaffiltext{1}{\footnotesize Graduate School of Chinese Academy of Sciences,
No. 19A, Yuquan Road, Beijing 100049,
China}

\begin{abstract}
We report the discovery of the likely orbital period of the ultracompact 
low-mass X-ray binary (LMXB) \ucxb.  Using time-resolved 
optical photometry carried out with the 8-m Gemini South Telescope, 
we obtained a 2.4-hr long, Sloan $r'$ light curve of \ucxb\ and
found a periodic, sinusoidal modulation at 17.4$\pm0.1$ min with a 
semiamplitude of 0.015$\pm0.002$ mag, which we identify as the binary period. 
In addition to 4U 0513$-$40 in the globular cluster NGC~1851 and 
the Galactic disk source 4U 1543$-$624, \ucxb\ is the third member of 
the ultracompact LMXBs that have orbital periods around 18 min.
Our result verifies the suggestion of \ucxb\ as an ultracompact 
binary based on its X-ray and optical spectroscopic properties. 
Given that the donor in \ucxb\ has been suggested to be either 
a C-O  or He white dwarf,
its likely mass and radius are around 0.024--0.029 $M_{\sun}$ 
and 0.03--0.032 $R_{\sun}$, respectively, for the former case 
and 0.034--0.039 $M_{\sun}$ and
0.033--0.035 $R_{\sun}$ for the latter case.
If the optical modulation arises from X-ray heating of the mass donor,
its sinusoidal shape suggests that the binary has a low inclination angle, 
probably around 10\arcdeg. 

\end{abstract}


\section{INTRODUCTION}

Low-mass X-ray binaries (LMXBs) constitute a large fraction of
bright X-ray sources ($L_{\rm X}\sim10^{36}$ erg s$^{-1}$) in the Galaxy. 
These binary
systems consist of an accreting compact star, either a neutron star 
or black hole, and a Roche lobe-filling, low-mass companion. 
Thus far approximately $\sim$200 LMXBs are known.
Among them, there is a class called ultracompact binaries.
Different from the majority of LMXBs which contain ordinary, hydrogen-rich
mass donors, ultracompact systems are believed to consist of
extremely low-mass, either hydrogen-poor or degenerate, companion stars 
\citep*{nrj86,ynv02}.
As a result, while ordinary LMXBs have a minimum orbital 
period around 80~min (\citealt{ps81}; \citealt*{rjw82}), ultracompact systems 
can evolve to extraordinarily small binary separations 
with orbital periods as short as a few minutes \citep*{prp02,nr03}.  
These ultracompact LMXBs, along with their white dwarf
analogues (the AM CVn binaries; see \citealt{war95}),  
represent extreme and exotic endpoints in
binary and stellar evolution.  

While the ultracompact systems had initially been assumed to be relatively
rare, the number known has more than doubled to 11 (including 4 globular
cluster sources) over the past few years,
with a range of orbital periods from 11 to 55 minutes \citep{ml09,zur+09}.
It is likely that there are more such binaries, because a few candidate 
systems have been identified either by their peculiar X-ray and/or optical 
spectral features (\citealt*{jpc01}; \citealt{nel+04}; \citealt{wan04}) 
or through their unusually low optical--to--X-ray flux ratios 
(\citealt*{dma00}; \citealt{bas+06}; \citealt*{ijm07}).
To fully study and understand the ultracompact LMXB population, verification of 
those candidate systems are warranted. It has shown that the indirect methods
for ultracompact binary identification may not be reliable \citep{swh07}. 
In order to verify their ultracompact
nature, time-resolved photometry for detecting orbital periodic signals 
is needed.
Moreover once ultra-short orbital periods are found, properties of 
the binary systems can be further estimated (e.g. \citealt{wc04}), helping our
understanding of these systems.
In an effort to verify the ultracompact nature of the proposed candidates, 
we have undertaken optical observations aiming to detect orbital flux
modulations. We have successfully found
the orbital period of the candidate 4U 1543$-$624 
\citep{wc04}. In this paper we report our discovery of the likely orbital 
period of another candidate \ucxb.

The LMXB \ucxb\ has been a bright X-ray source 
($L_{\rm X}\sim 10^{36}$ erg s$^{-1}$) 
and detected by all major X-ray satellites (\citealt{jc03} and
references therein). 
On the basis of comparison of its X-ray spectrum to that of the known
ultracompact LMXB 4U 1626$-$67, the source has been suggested to be
an ultracompact binary with a neon-enriched degenerate donor 
\citep{jpc01,jc03}. Probably because the binary has a low 
inclination angle (generally $i<$ 60\arcdeg; \citealt*{fkr02}),
no orbital signals were found in X-ray observations of the source
\citep{jc03}. 
The optical counterpart to \ucxb\ was identified by 
\citet{ci87}, $V = 21$, $B-V = 0.3$.  
Based on its optical--to--X-ray flux ratio, the orbital period of the binary 
has been suggested to be $\lesssim 60$ min \citep{jc03}. The source distance
is probably 4.1--5.4 kpc, estimated from type-I X-ray bursts detected 
from the source \citep{iz+05}.

\section{OBSERVATIONS AND DATA REDUCTION}    

Time-resolved imaging of \ucxb\  was carried out on 2008
December 5 using the 8-m Gemini South Telescope. The instrument was
Gemini Multi-Object Spectrograph (GMOS; \citealt{hoo+04}), whose detector
array consists of three 2048$\times$4608 EEV CCDs. We used only the middle
CCD chip (CCD~02) for imaging. The pixel scale 
is 0.073\arcsec/pixel, while the detector was 2$\times 2$ binned
for our observation. A Sloan $r'$ filter with the central wavelength
at 6300 \AA\  was used. We obtained 179 continuous frames with 
an exposure time of approximately 24.5 sec. Including 24 sec readout time
for each exposure, the total length of the observation was 
approximately 2.4 hrs.
The observing conditions during the early part of our observation 
(first 59 frames) were relatively poor, with the average seeing [FWHM of
the point-spread function (PSF) of the images] being $\simeq 1.0\arcsec$
and a few frames having as large as 1.3\arcsec\ --1.5\arcsec\ seeing. For
the remaining part, the conditions were good with most of the frames
having $\lesssim 0.8\arcsec$ seeing.
\begin{center}
\includegraphics[scale=0.65]{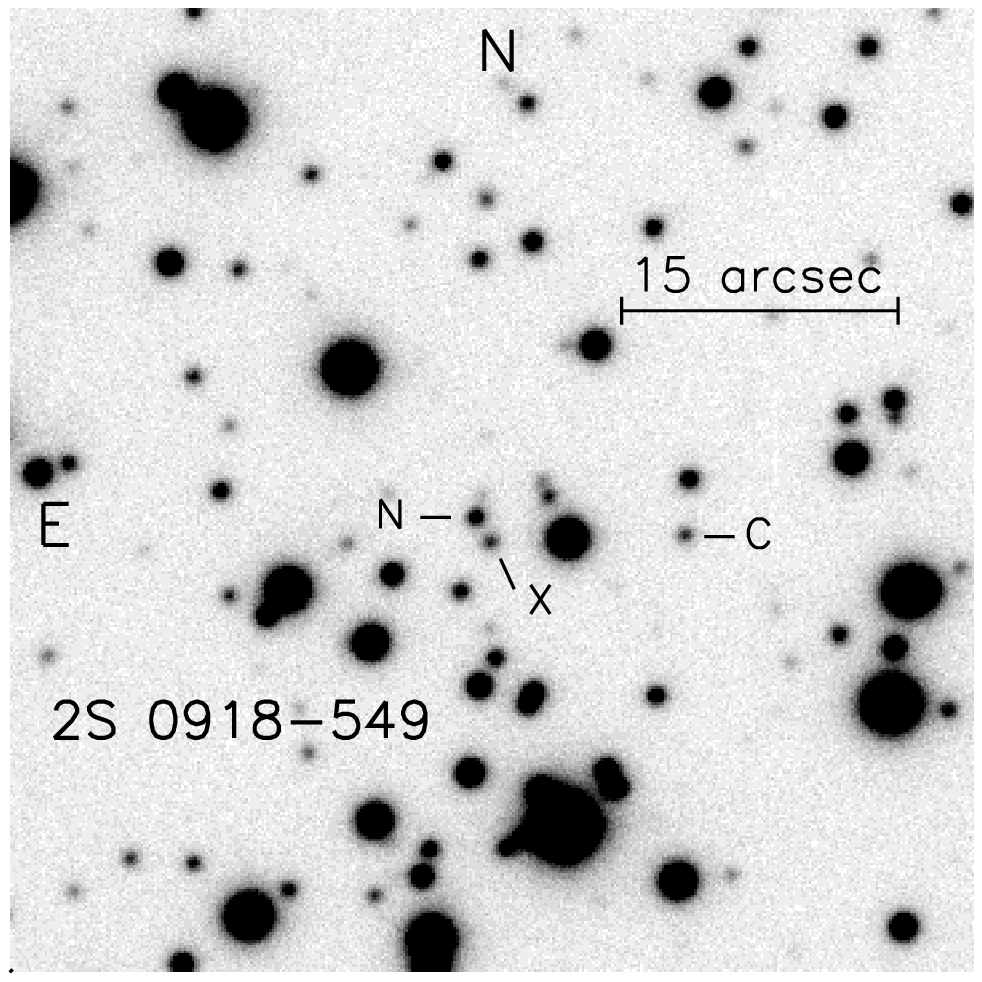}
\figcaption{Gemini South $r'$ image of the \ucxb\  field. 
Object $X$ is the optical counterpart to \ucxb.  The nearby star
labeled as $N$ is 1.5\arcsec\ away from the target. 
The star labeled as $C$ was used as a check star. 
\label{fig:finding} }
\end{center}

We used the IRAF packages for data reduction. The images were bias subtracted
and flat fielded. The bias and flat frames were from GMOS baseline 
calibrations taken during the same night. 

We performed PSF-fitting photometry to obtain brightnesses of 
our target and other in-field stars, with a photometry program {\tt DOPHOT} 
\citep*{sms93} used. A finding chart 
of the target field is
shown in Figure~\ref{fig:finding}. As identified by \citet{ci87}, 
there is a nearby star (labeled as $N$) 1.5\arcsec away from our target. 
To avoid possible 
contamination from this nearby star caused by the poor observing conditions
during the early part of our observation, 
we positionally calibrated the first 59 frames to a reference image that
was combined from three high-quality frames. The positions
of star $N$ and \ucxb\ were determined in the reference image, and were fixed
at the positions for photometry of the first 59 frames. At last
we excluded three frames among them from the data.  
From the three frames, the brightness measurements of star $N$ and other 
in-field stars obtained were not consistent with the average brightnesses 
from other frames, and we note that the three frames have the seeing of 
1.3\arcsec--1.5\arcsec. 

Differential photometry was performed to eliminate systematic flux variations
in the images. Three isolated, nonvariable bright stars in the field were
used. The brightnesses of our targets and other stars in each frame were
calculated relative to the total counts of the three stars. A field star
labeled as $C$ (see Figure~\ref{fig:finding}) was used as a check star, 
as it was nonvariable and had similar brightness to our target.

Because we did not request observations of standard stars for flux calibration
in our Gemini program and also no standard stars were imaged
in the same filter by the Gemini South
telescope within at least half a month before and after our observation, 
we used the $BV$ magnitudes of star $N$ measured by \citet{ci87} to obtain 
absolute magnitudes of the target and other stars. The transformation formula
between $r'$ and $BV$ magnitudes given by \citet{fuk+96} was used.
We found an average $r'$ magnitude of 20.95$\pm$0.16 for \ucxb, where the 
uncertainty comes from the relatively large uncertainties on the $BV$ magnitudes
of star $N$ in \citet{ci87}. We note that with the same transformation, 
$r'=20.94$ for \ucxb\ in \citet{ci87}, which indicates that the binary 
has not had significant
changes in its optical brightness. The average $r'$ magnitude of star $C$ was
20.98$\pm 0.16\pm 0.02$, where 0.02 mag is the standard deviation of star
$C$ measured from 176 frames.

\section{RESULTS} 

\begin{figure*}
\begin{center}
\includegraphics[scale=0.8]{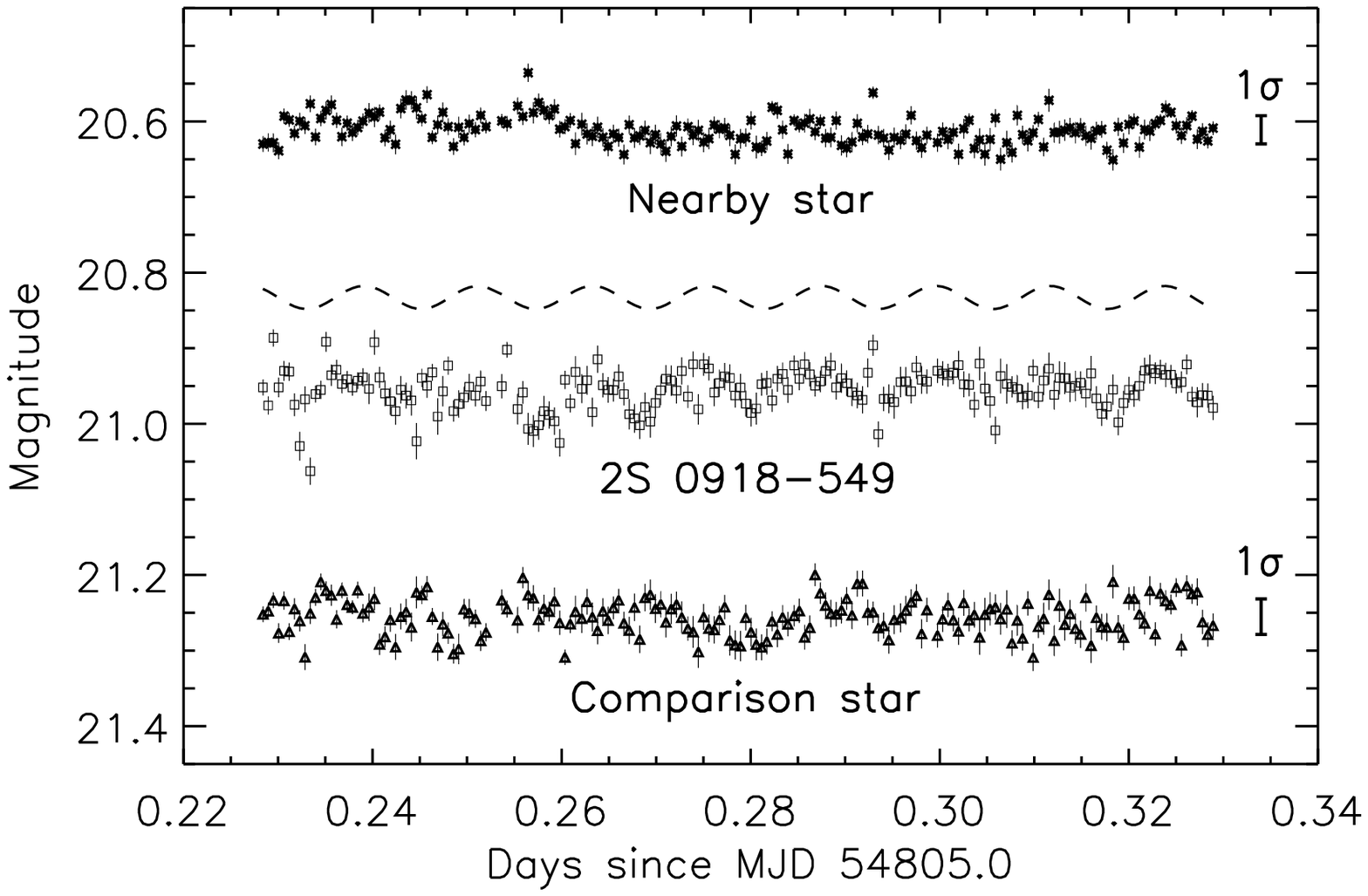}
\figcaption{$r'$ light curve of \ucxb\ (squares). The light curves of 
the nearby star $N$ (asterisks) and comparison star $C$ (triangles),
down-shifted by 0.34 and 0.28 mag respectively, are also shown. 
A sinusoid (dashed curve) is plotted to help indicate the periodic 
modulation detected in the light curve of \ucxb.
\label{fig:lc}
}
\end{center}
\end{figure*}

In Figure~\ref{fig:lc} the obtained light curves 
of \ucxb, star $N$, and the check star
$C$ are shown. A periodic modulation in the light curve of \ucxb, 
while with a low-amplitude, is clearly visible. 
To determine its period,
we applied a phase-dispersion minimization technique \citep{ste78} with
16 bins of the full phase interval (0, 1) used. The resulting periodogram
is shown in Figure~\ref{fig:pdm}. The $\Theta$ statistic indicates the
detection of a periodicity and its two harmonics. Fitting the region
near the first minimum with a parabola \citep{ste78}, we found period
$P=17.4$ min.

In order to quantify the overall periodic modulation in the light curve,
we fit the light curve with a sinusoid. The best-fit has reduced $\chi^2=2.3$
for 172 degrees of freedom (DOF), and from the best-fit we found 
$P=17.38\pm0.13$ min and a semiamplitude of 0.014$\pm$0.002 mag. The large
$\chi^2$ value is mainly caused by large scattering of the first 56 data 
points due to the poor observing conditions. Excluding them and fitting 
the remaining data points, we found
reduced $\chi^2=1.2$ for 117 DOF ($P$ was fixed at 17.4 min). The obtained
semiamplitude was 0.015$\pm$0.002 mag, not having significant changes.
Therefore the modulation can be described by a sinusoid with a semiamplitude
of 0.015 mag. The folded light curve at $P=17.4$ min as well as the best-fit
sinusoid are shown in Figure~\ref{fig:folded}.
The time at the maximum of the sinusoidal fit (phase zero)
was MJD 54805.23281$\pm$0.00027 (TDB) at the solar system barycenter.

\section{DISCUSSION}

Using Gemini high time-resolution imaging we obtained an accurate optical 
light curve of \ucxb\ and have discovered a periodic flux 
modulation in the light curve.  A low-amplitude modulation is clearly visible 
and appears
to be coherent. Given the known X-ray and optical properties of \ucxb,
it is very likely that we have verified the ultracompact nature of this binary
and its orbital period is around $17.4$ min (see discussion below). In addition 
to 4U 0513$-$40 in the globular cluster NGC~1851 \citep{zur+09} 
and the Galactic disk source 4U 1543$-$624 \citep{wc04}, \ucxb\ is the third 
member of the ultracompact binaries that have orbital periods around 18 min.
\begin{center}
\plotone{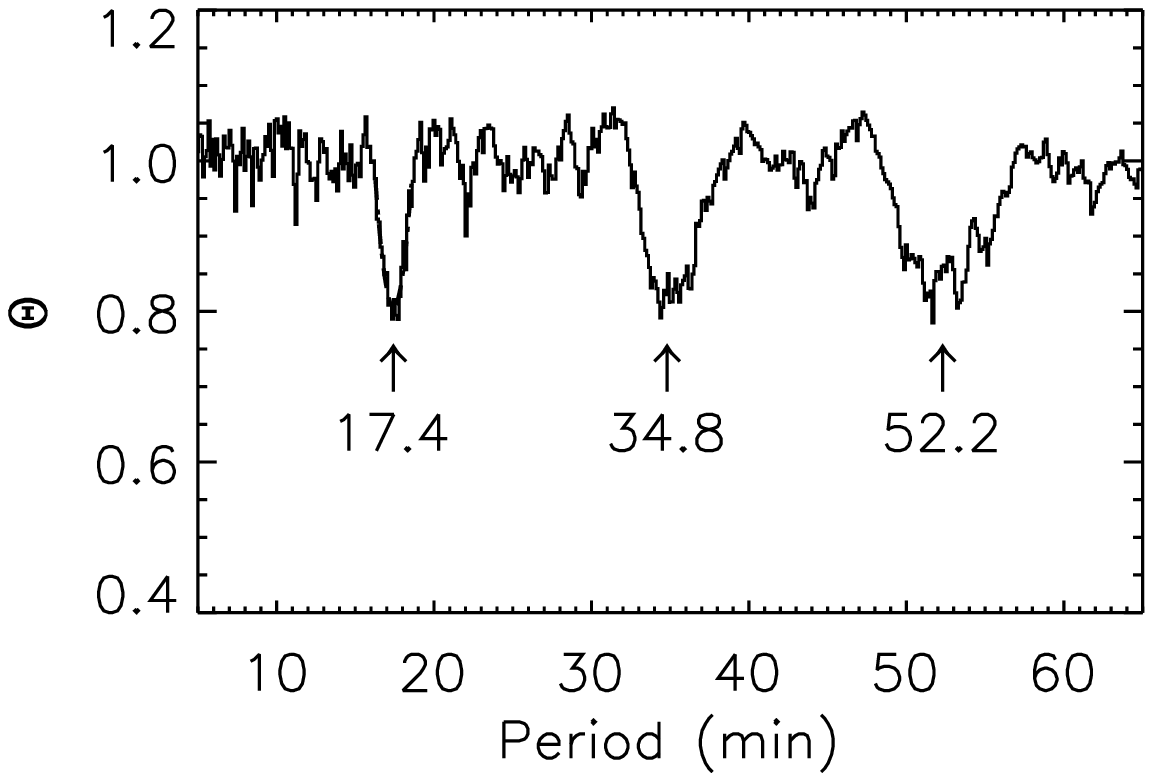}
\figcaption{Phase-dispersion minimization periodogram. 
The positions of the minimum $\Theta$ statistic at 17.4 min and its two
harmonics are indicated by arrows.
\label{fig:pdm}
}
\end{center} 

We use the discovered orbital period to estimate the mass 
and radius of the donor. Since the mean density of a Roche lobe-filling 
companion is determined by the binary period, our 17 minute period defines 
a mass-radius relation for the companion, shown as the solid curve in 
Figure~\ref{fig:mr}. 
On the basis of the high Ne/O abundance ratio measured through X-ray 
spectroscopy, the donor in \ucxb\  has initially been suggested to 
be a low-mass C-O white dwarf \citep{jpc01, jc03}. However, the analysis 
of several type-I X-ray bursts detected from the source possibly indicates 
that the donor instead is a helium white dwarf \citep{iz+05}. 
In any case to compare the donor to stellar models, 
we use the M-R relations for different types of white dwarfs 
provided by \citet{db03}. Because extremely low-mass white dwarf donors 
in ultracompact systems may be thermally bloated compared to cold stars, 
affecting their M-R relation \citep{bil02,db03},  we show both cold and 
hot solutions for pure He, C, and O white dwarfs in Figure~\ref{fig:mr}.
A helium white dwarf with a mass of 0.034--0.039 $M_{\sun}$ and a radius 
of 0.033--0.035 $R_{\sun}$, or a C/O white dwarf with a mass of 
0.024--0.029 $M_{\sun}$ and a radius of 0.03--0.032 $R_{\sun}$ 
can fit in the Roche lobe-filling donor.

Modulation of an optical light curve for LMXBs generally arises from  
the companion star that is heated by the central X-ray source, with the visible 
area of the heated face varying as a function of orbital phase and the superior
conjunction of the companion star corresponding to the observed brightness 
maximum of the light curve (e.g., \citealt{vm95}). It has also been realized
that compact LMXBs with extreme mass ratios (such as ultracompact binaries)
are potential superhump sources \citep{has+01}. The variation in the light
curve of a superhump binary arises from an elliptical accretion disk, 
which is developed when the disk extends beyond the 3:1 resonance radius
and precesses in the inertial frame due to the tidal force of a secondary star 
(e.g., \citealt{wk91}). Without an independent determination of the
binary period (see, e.g., \citealt{wc10}), we can not distinguish between
the two possibilities for the modulation seen in \ucxb. A superhump modulation
may have an asymmetric shape (e.g., \citealt{wc10}). However in the current
light curve we obtained, no asymmetry is clearly seen.
In either case since superhump periods are only
a few percent longer than the corresponding orbital periods, the orbital period
of \ucxb\ is around 17.4 min. 
\begin{center}
\includegraphics[scale=0.7]{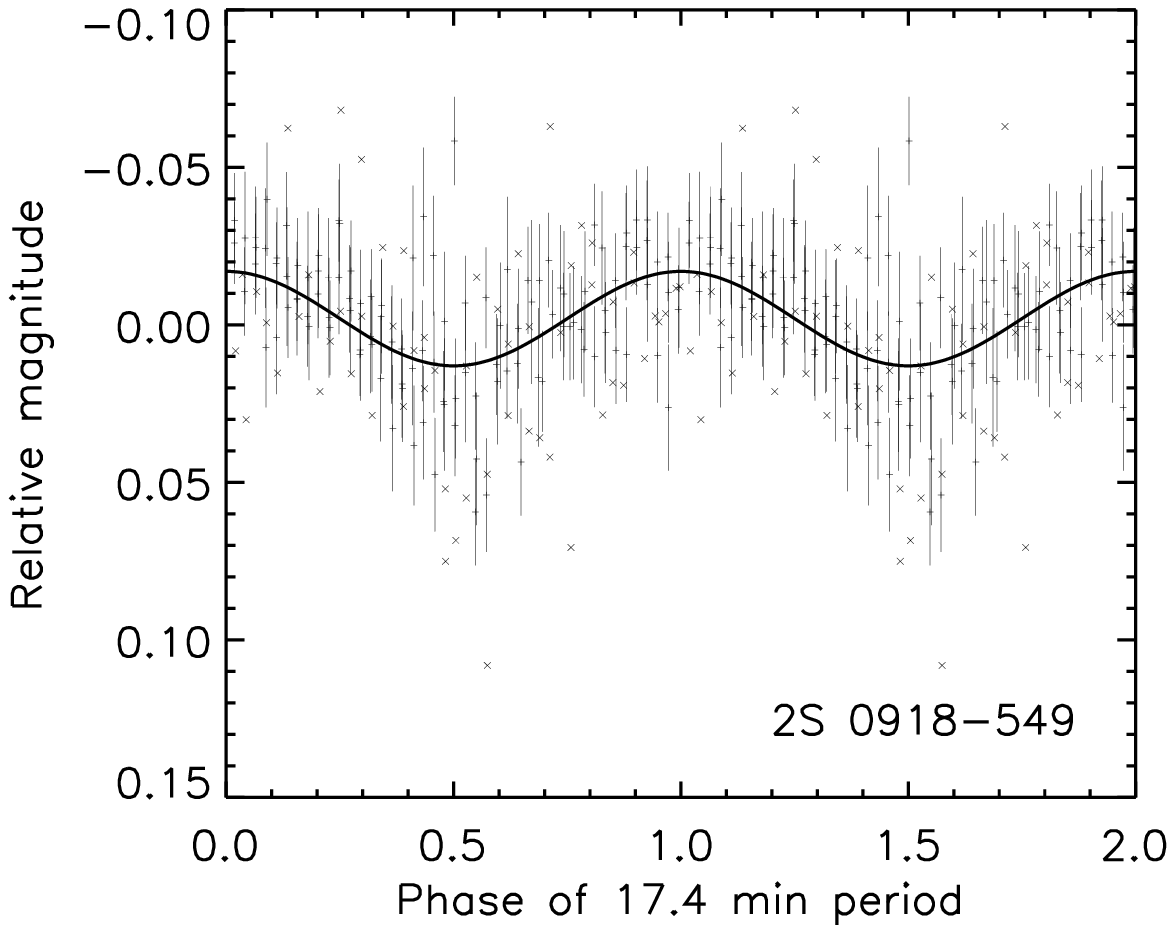}
\figcaption{$r'$ light curve of \ucxb\ folded at 17.4 min. 
Two cycles are displayed for clarity.
The solid curve indicates the best-fit sinusoid with a semiamplitude of
0.015$\pm$0.002 mag. The first 56 data points that were excluded
from the fit are shown as crosses. 
\label{fig:folded}
}
\end{center}

Considering that the periodic modulation arises from the companion star, 
the inner face of the companion star in \ucxb\ is heated by X-ray emission from
the central neutron star and its effective temperature can be estimated.
The 0.1--200 keV X-ray luminosity 
$L_{\rm X}$ is 1.9$\times 10^{36}\ d_5^2$ erg s$^{-1}$, 
where $d_5$ is the source distance assumed to be 5 kpc and the unabsorbed 
X-ray flux $F_{\rm X}= 6.4\times 10^{-10}$ erg s$^{-1}$ cm$^{-2}$ given 
by \citet{iz+05} is used.
The fraction $f$ of the X-ray energy absorbed by the companion 
is $f=\eta_{\ast}(R_2/D_b)^2/4
\simeq$0.004$\eta_{\ast}(R_2/0.032\ R_{\sun})^2 [1+(q/0.021)]^{-2/3}$, 
where $\eta_{\ast}\sim 0.5$ is the fraction of the received X-ray energy 
absorbed by the companion, $R_2$ is the radius of the companion,  
and $D_b$ is the binary separation distance 
($D_b\sim 1.7\times 10^{10}$ cm for orbital period $P_{\rm orb}=17.4$ min). 
The mass ratio $q=M_2/M_{\rm ns}$,
where $M_{\rm ns}$ is the neutron star mass and we assume 
$M_2=0.03\ M_{\sun}$ and $M_{\rm ns}=1.4\ M_{\sun}$. 
Following \citet{ak93}, the effective temperature of the companion's inner face 
is $T=(fL_{\rm X}/\pi R_2^2\sigma)^{1/4}\simeq 46000 d_5^{1/2}$ K, 
where $\sigma$ is the Stefan-Boltzmann constant. 
The visible area of this hot face varies as a function of the orbital phase,
yielding a modulation of $[1+\sin i\sin (2\pi t/P_{\rm orb})]$,
where $i$ is the inclination angle of the binary
(see details in \citealt{ak93}).
Using such a modulation function,
we can test how the observed modulation is generated.
The extinction to the source $A_V\simeq 1.65$,
estimated from $A_V=N_{\rm H}/1.79\times 10^{21}$~cm$^{-2}$ \citep{ps95}
by using hydrogen column density to the source 
$N_{\rm H}=2.95\times 10^{21}$ cm$^{-2}$ \citep{jc03}.
By adding a constant flux component (arising from the accretion disk)
to the modulation function and fitting
the dereddend light curve of \ucxb\ with the modulation function, we find 
$i\sim 10\arcdeg$. 
The estimated low inclination angle is consistent with
the non-detection of orbital signals at X-ray energies, 
which generally indicates $i\lesssim 60\arcdeg$ (\citealt{fkr02}), 
and likely explains the low-amplitude modulation in the light curve.
In order to fully explore properties of the binary
by fitting the optical modulation, an advanced binary light curve
model is needed (e.g., \citealt{del+08}; Wang et al. 2011, in preparation).
\begin{center}
\includegraphics[scale=0.6]{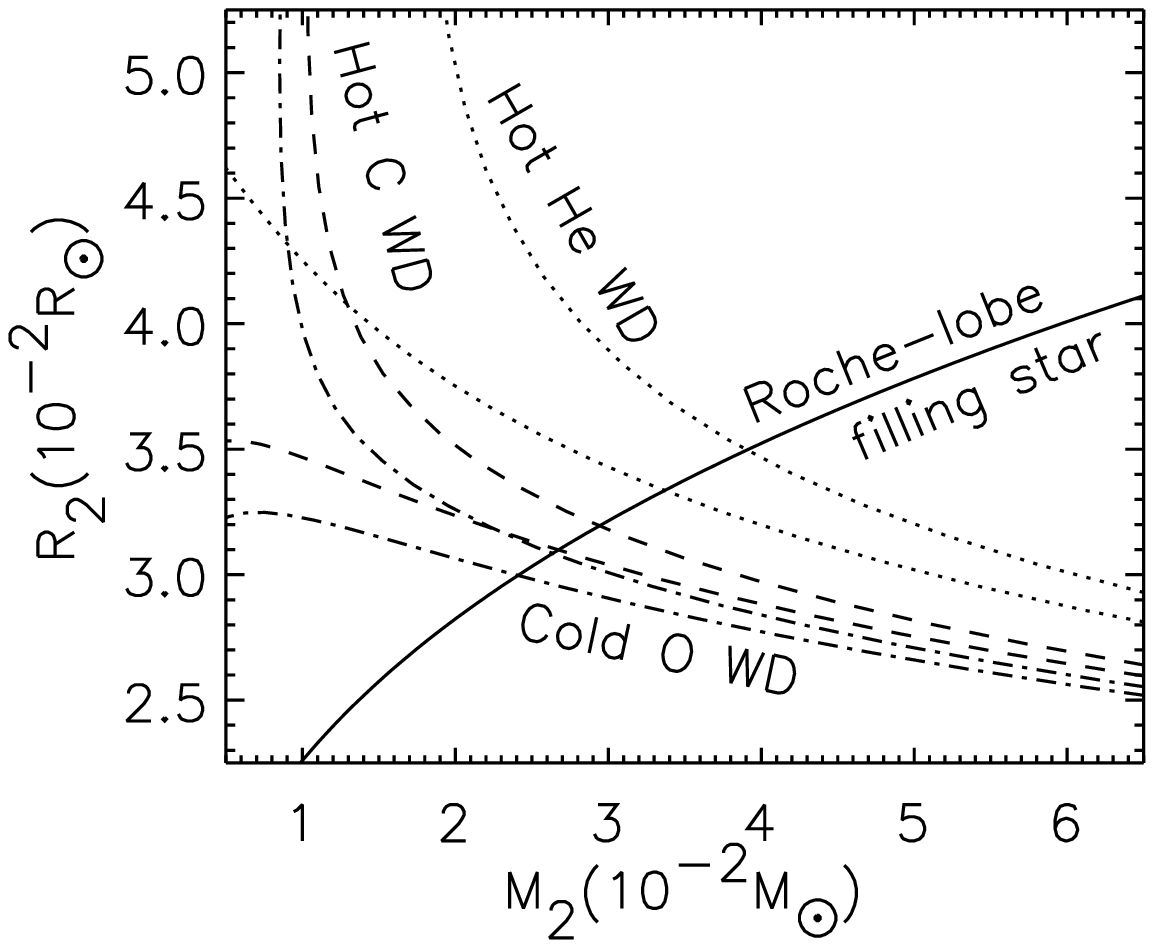}
\figcaption{Mass and radius values constrained for the companion star 
in \ucxb. The solid curve is the mass-radius relation for a Roche lobe-filling 
donor in a 17.4 min binary. The dotted, dashed, and dash-dot curves represent
the model curves for low-mass He, C, and O white dwarfs from \citet{db03},
respectively. For each type of white dwarf, both cold (10$^{4}$ K) and 
hot (3 $\times 10^{6}$ K) core temperature model curves are shown.
\label{fig:mr}
}
\end{center}

As a separate check, we also estimate the distance to \ucxb.
Mass transfer in ultracompact binaries is driven by gravitational radiation, 
and a mass transfer rate in \ucxb\ can be estimated to be
\[
\dot{M}\simeq 6.2\times 10^{-10} M_{\sun}\ {\rm yr}^{-1} \left(\frac{M_{\rm ns}}
	{1.4\ M_{\sun}}\right)^{2/3}\left(\frac{M_2}{0.03\ M_{\sun}}\right)^2 
\]
\[
	\times\left(\frac{P_{\rm orb}}{17.4\ {\rm min}}\right)^{-8/3}\ \ \ .
\]
Since $L_{\rm X}= GM_{\rm ns}\dot{M}/R_{\ast}=4\pi d^2F_{\rm X}$, 
the unabsorbed 0.1--200 keV X-ray flux of \ucxb\ would imply 
$d\sim$9 kpc, where conservative mass transfer onto a 1.4 $M_{\sun}$ neutron 
star is assumed and $R_{\ast}\simeq 10$ km is the neutron star radius.
The distance value is larger than the 4.1--5.4 kpc range derived 
from type-I X-ray bursts. We note that $\dot{M}$ is sensitive to $M_2$. For
example, if $M_2=0.024\ M_{\sun}$, where an oxygen white
dwarf has to be assumed, the distance could be lowered to 7 kpc.
A larger X-ray flux can also help lowered the distance.
As recorded in an $ASCA$ X-ray observation in 1995, approximately 7 times 
larger X-ray flux was detected \citep{jc03, iz+05}. If that $ASCA$ flux 
is considered, the distance would be lowered to $\sim$4 kpc.
However, given the known X-ray flux history of \ucxb, its 2--10 keV flux
has been stable and around $10^{-10}$ erg s$^{-1}$ cm$^{-2}$ \citep{iz+05}, 
suggesting that the $ASCA$ flux was only a one-time event.

\acknowledgements{The Gemini queue mode observation was carried out under 
the program GS-2008B-Q-78.
The Gemini Observatory is operated by the Association of Universities 
for Research in Astronomy, Inc., under a cooperative agreement with 
the NSF on behalf of the Gemini partnership: the National Science Foundation 
(United States), the Science and Technology Facilities Council 
(United Kingdom), the National Research Council (Canada), CONICYT (Chile), 
the Australian Research Council (Australia), CNPq (Brazil), and CONICET 
(Argentina). This research was supported by the starting funds of Shanghai
Astronomical Observatory, National Natural Science
Foundation of China (11073042), and National Basic Research Program of China
(973 Project 2009CB824800). Z.W. is a Research Fellow of the 
One-Hundred-Talents project of Chinese Academy of Sciences.
}

{\it Facility:} \facility{Gemini:South (GMOS)}


\end{document}